\documentclass[apj]{emulateapj}
\usepackage{mathptmx}
\usepackage{listings}

\def\gtorder{\mathrel{\raise.3ex\hbox{$>$}\mkern-14mu
             \lower0.6ex\hbox{$\sim$}}}
\def\ltorder{\mathrel{\raise.3ex\hbox{$<$}\mkern-14mu
             \lower0.6ex\hbox{$\sim$}}}

\slugcomment{Draft of \today}

\shorttitle{Astrometry}

\shortauthors{Ofek}

\begin{document}

\title{A code for robust astrometric solution of astronomical images}
\author{
E.~O.~Ofek\altaffilmark{1},
}
\altaffiltext{1}{Benoziyo Center for Astrophysics, Weizmann Institute
  of Science, 76100 Rehovot, Israel.}

\begin{abstract}

I present a software tool for solving the astrometry of astronomical images.
The code puts emphasis on robustness against failures
for correctly matching the sources in the image to a reference catalog,
and on the stability of the solutions over the field of view
(e.g., using orthogonal polynomials for the fitted transformation).
The code was tested on over $5\times10^{4}$ images
from various sources, including the Palomar Transient Factory (PTF)
and the Zwicky Transient Facility (ZTF).
The tested images equally represent low and high Galactic latitude
fields and exhibit failure/bad-solution rate of $\ltorder2\times10^{-5}$.
Running on PTF 60-s integration images, and using the GAIA-DR2 as a
reference catalog,
the typical two-axes-combined astrometric root-mean square (RMS)
is 14\,mas at the bright end,
presumably due to astrometric scintillation noise and systematic errors.
I discuss the effects of seeing, airmass and the order of the transformation
on the astrometric accuracy.
The software, available online, is developed in {\tt MATLAB} 
as part of an astronomical image processing environment and it can be run
also as a stand-alone code.

\end{abstract}

\keywords{
techniques: image processing -- methods: data analysis}

\section{Introduction}
\label{Introduction}

There is a large variety of existing software for astrometric solutions
(e.g., SCamp, Bertin 2006, 2010;
Astrometry.net, Lang et al. 2010, 2012;
ASCfit, J{\o}rgensen et al. 2002, Pickles 2012).
Some of these tools are extremely powerful.
For example, {\tt astrometry.net} is capable of matching star
patterns in an image-catalog
to a reference-catalog, without prior knowledge on the image's sky position,
rotation or scale.
This feature is valuable, for example, when solving the astrometry
of ancient photographic plates for which the plate scale,
rotation and sky location is errornous or unavailable (e.g., Tang et al. 2013).
However, in some cases, the approximate sky position,
plate scale and rotation are known,
and simpler, quicker and more robust approaches can be used (e.g., Bertin 2006).

Here I present a software tool for solving the astrometry of
astronomical images, particularly wide-field and dense fields.
This code attempts to improve over some existing tools in one aspect,
which is robustness
against failures and inaccurate solutions.
My main motivation was based on the Palomar Transient Factory
(PTF; Law et al. 2009; Rau et al. 2009)
pipeline (Laher et al. 2014), which
uses SCamp (Bertin 2006) and Astrometry.net (Lang et al. 2010).
A total of about 4\% of the PTF images are reported to have poor
astrometric solutions (see Laher et al. 2014),
and this fraction can reach about 30\% in some fields
near the Galactic plane.
In many cases, this is due to anomalous distortion coefficients,
or poor solutions in some parts of the image.
We identify two main problems:
The first issue is that in some cases the correct solution cannot be found.
This may happen, for example, in fields with high stellar density,
or in large field-of-view images taken at high airmass,
which causes the plate scale to be position-dependent.
The second issue is that, sometimes, fitting a high order polynomial transformation
returns an adequate solution in some region of the field but provides
a poor fit in other regions.
This specific problem may happen, for example,
when the fitted transformation is ill-conditioned.

Our strategy to mitigate these two problems is relatively straightforward.
The solution consists of two minor modifications
to the typical approach, which can be implemented in other tools.
The first improvement is to divide the image into several smaller sub-images,
and then attempt to mach the stars to the catalog sources
in each sub-image separately.
In this case it is easy to identify and reject bad solutions
(i.e., a solution that is different from
the solutions of the other sub images).
Moreover, the good solutions can be used to interpolate over the
regions with bad solutions.
The second strategy is to use orthogonal polynomials for the transformation,
or transformations that have a physical motivation (e.g., tip-tilt
transformation).
A feature of the current version of the
code over, e.g., Astrometry.net, is that
it requires rough knowledge of the image coordinates and plate scale
and hence needs to be configured for each telescope/instrument
on which it is used.

I describe the algorithm's main steps in \S\ref{sec:description}.
In \S\ref{sec:Per}, I discuss the code performance,
while the code architecture is described in \S\ref{sec:code}.
I summarize in \S\ref{sec:Sum}.

\section{General description of the code}
\label{sec:description}

The algorithm starts with either the image or catalog of sources
in the image.
If the image is provided without a catalog, the code
finds the sources in the image and measures their X/Y positions
and instrumental magnitudes using {\tt mextractor} (Ofek et al. in prep)
or {\tt SExtractor} (Bertin \& Arnouts 1996).
The code can use a mask image, and a
bitmask-dictionary\footnote{A mask-dictionary relates a problem to a bit-index.},
that indicate problems per pixel in the input image.
The mask image can be used to remove sources that may badly
affect the solution (e.g., cosmic rays, saturated pixels).
Furthermore, the code can remove sources based on a surface density criteria.
This is done by identifying regions with a large number of sources per unit area,
and rows/columns in which the number of sources is above the mean source row/column density.
This step is important as very high density regions can sometimes be
matched with a wrong solution.
The user supplies the code with approximate image coordinates,
radius in which the reference catalog sources are
extracted\footnote{Due to uncertainties
in the image position, the catalog search radius can be larger than the image size},
possible coordinate flips,
plate scale, 
and, optionally, the range of the image rotation
(i.e., the position angle of the Y-axis).

Next, the code extracts the stars from the reference catalog
according to user- or header-specified approximate coordinates.
The code can access a large number of online reference catalogs
(e.g., VizieR, Ochsenbein et al. 2000),
user-supplied catalogs, or
local catalogs in the {\tt catsHTM} format (Soumagnac \& Ofek 2018)
that provide fast access to large catalogs.
If the reference catalog includes proper motions
(and, optionally, parallaxes and radial velocities),
the coordinates are transformed to the image-catalog epoch.
The catalog coordinates are then projected onto a plane
using the gnomonic (tangential) projection (e.g., Calabretta \& Greisen 2002)
and the known image plate scale.
If needed, the code may equalize the source surface density
in the image catalog and reference catalog by removing the faintest sources
in the image-catalog or reference-catalog -- whichever contains more sources
per unit area.

Next, the code attempts to match the sources in the image catalog
to the projected reference-catalog sources.
The image catalog is partitioned into sub catalogs
of smaller regions of the image
(i.e., sub-images with default size of $1024\times1024$\,pixels),
and the matching process
is done separately for each region.
This step is designed to make the code robust against plate-scale changes
as a function of position
and to deal with failures to match the catalogs -- i.e.,
it increases the probability that at least some of the stars
in the field will be matched to the reference catalog.
By default, the code first attempts to find possible rotational solutions
using the algorithm suggested by Kaiser et al. (1999)
and used in {\tt SCamp} (Bertin 2006).
This is done by calculating the distances and position angles
between all the pairs of sources in the image catalog,
and all the pairs of sources in the reference catalog.
Next, a two dimensional histogram of distance
and position angle is generated for each catalog,
and the two histograms are cross-correlated to find
possible rotations between the image catalog and reference catalog.
Alternatively, if this step fails or the user
chooses to skip this step, the code can look for possible shifts,
between the catalogs,
for a range of possible rotations specified by the user.

Following the algorithm implemented by Giveon et al. (1999),
for each possible rotation and coordinate flip,
the code calculates
all the possible shifts in X and Y coordinates
between the sources in the image and the projected sources
in the reference catalog.
A two-dimensional histogram (in X and Y) of all the possible
differences is calculated and peaks in the histogram are identified.
By default, the code does~not select just the highest peak in each histogram,
rather, it includes all the peaks satisfying some
user-specified criteria (e.g., signal-to-noise ratio).
The result is that the code collects many possible solutions
with different rotations, flips, and shifts.
Next, each one of the possible solutions is tested.
This is done, for each candidate solution, by fitting an affine
transformation between the
image catalog sources and reference catalog sources.
For each transformation, the root-mean square (RMS) of residuals and
the number of matched stars are reported,
and the best solution is selected based on some user-specified criteria
(e.g., RMS and/or number of matched stars).

The code compares
the solutions of the various sub-images and reject sub-images with
solutions which
are different than the median solution.
Therefore, this step may result in sub-images without matched stars.
The code uses the matched stars over all
sub-images to fit a general transformation between
the image catalog and reference catalog.
The available transformations include
affine transformation, tip-tilt terms (e.g.,
$\alpha X(X+Y)+\beta X(X-Y)$, where $\alpha$ and $\beta$ are free parameters;
van~Altena 2012),
general polynomials, and orthogonal Chebyshev polynomials.
The default transformation depends on the number of matched sources.
The fit includes iterative removal of outliers.
The default is to use two iterations with sigma clipping
of $5$-$\sigma$ after the first iteration.
In the second iteration a weighted fitting is implemented,
where the weights are the inverse variance.
Here, the variances are obtained
from the RMS as a function of magnitude plot of the first-iteration fit.
Note that a more aggressive sigma-clipping may lead,
in some cases, to the rejection of good reference stars
and bias in the estimated transformation.

The transformation parameters are converted into
a polynomial representation and added to the image header in the
TPV World Coordinate System representation\footnote{https://fits.gsfc.nasa.gov/registry/tpvwcs/tpv.html} (see also Calabretta \& Greisen 2002).

\section{Performances}
\label{sec:Per}

I tested the code on over 50,000 images from various sources.
The main data set used is from the Palomar Transient Factory
(PTF; Law et al. 2009; Rau et al. 2009).
The code was also tested on images from
the Zwicky Transient Facility (Bellm et al. 2018),
the Palomar $200''$ WIRC infrared camera,
and the 28$''$ telescope at the Wise observatory.
The PTF data reduction is described in Laher et al. (2014)
and Ofek et al. (2012).
Each PTF image\footnote{The PTF camera has 11 active CCDs and we refer to the data from each CCD as an image.}
is 2048$\times$4096 pixels,
with a $1.01''$\,pix$^{-1}$ scale.
The ZTF data reduction pipeline is described in Masci et al. (2018).
The ZTF images\footnote{The ZTF camera has 16 CCDs, each with four amplifiers. The data from each amplifier is treated as an individual image.}
are 3080$\times$3072 pixels, with the same pixel scale
as in the PTF images.

Testing the code on over 50,000 images, I encountered zero failures.
Here failure is defined as either inability to find a solution,
or solutions with large RMS  in some regions of the image (see details below).
In cases in which the number of matched stars was low (sometimes indicative
of a problem), I inspected the images by eye and verified the
solution was correct.

Below I present the statistics of the results from a homogeneous dataset
of 8,548 PTF images.
The images in this dataset were selected randomally by sky position,
so a large fraction of the images were
at low Galactic latitudes.

For the reference catalog, as a default, I used the GAIA-DR2 catalog
(Gaia Collaboration, et al. 2018).
However, the code was also tested using the GAIA-DR1 catalog
(Gaia Collaboration, et al. 2016)
and the UCAC-4 catalog (Zacharias et al. 2013).
For the transformation between the reference catalog
and the image-catalog, the code fitted
an affine transformation plus
a 4th order Chebyshev polynomial of the second kind.
In the fitting procedure, two iterations were performed.
Before the second iteration the code removed sources
using a 5-$\sigma$ clipping,
where the standard deviation, $\sigma$, was calculated as a function of magnitude.
Sources were extracted from the images using 
{\tt mextractor} (Ofek et al. in prep.).
I also tested the code on source catalogs
generated using {\tt SExtractor} (Bertin \& Arnouts 1996)
and {\tt DAOPHOT} (Stetson 1987).
Specifically, the {\tt DAOPHOT} catalogs
were generated by the PTF pipeline.
However, for the {\tt DAOPHOT} catalogs, I encountered
failures in about 1\% of the cases, due to some
issues with the PTF-generated catalogs.
Problems included invalid magnitudes and errors
in the Point-Spread Function (PSF) extraction.

\begin{figure}
\centerline{\includegraphics[width=8cm]{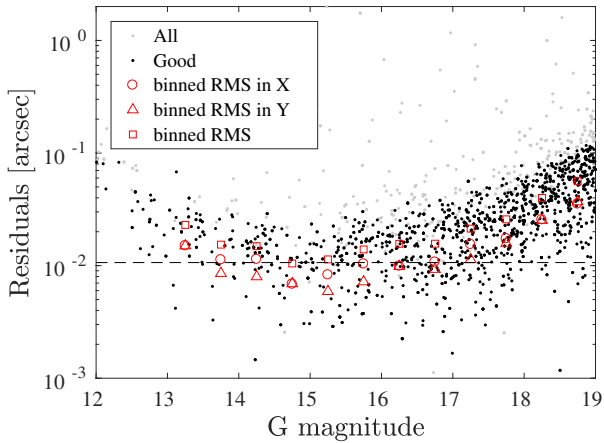}}
\caption{Two-axes combined Residuals vs. GAIA G-band magnitude of a PTF $R$-band image
taken on 2012 Jan 2.3661.
The black points show the sources used in the astrometric solution.
The gray points show the sources that were clipped
(e.g., due to mismatches, confusion, or extended sources).
The dashed line marks the level of the
measured asymptotic RMS (11\,mas).
The red circles, triangles, and squares show
the mean absolute value of the residuals in X axis, Y axis
(after sigma clipping),
and total residuals in 0.5\,mag bins, respectively.
Stars brighter than $\approx14.5$\,mag are saturated.
\label{fig:PTF_Mag_RMS}}
\end{figure}

Figure~\ref{fig:PTF_Mag_RMS} shows the residuals
(square root of sum of squares of $X$ and $Y$ residuals)
as a function of the GAIA magnitude.
This is shown for the first image in the dataset.
The minimum RMS, of about 11\,mas, is achieved for the brightest non-saturated
stars (i.e., magnitude $\approx15$).
Unless explictly indicated, all rms values are for the two-axes combined rms.
The single-axis rms is typically $\sqrt{2}$ times smaller than the two-axes combined rms.
At the bright end of the non saturated stars,
the solution is presumably limited
by atmospheric scintillation noise
and systematic errors, while at the faint-end,
the solution degrades and is limited
by Poisson noise.
To characterize the quality of the astrometric solution
we are interested in a metric which does~not depend on
the magnitude-range of stars used in the solution.
Therefore, the adopted estimator for the quality of the solution
is the RMS at the bright end for unsaturated stars.
For ground based, seeing-limited observations,
the RMS measured at the minimum of the magnitude-RMS plot,
is typically dominated by the scintillation noise and any additional systematic errors.
For this reason, I define the asymptotic RMS
as the minimum RMS of a second-order polynomial
fitted to the RMS vs. magnitude plot,
in the range of available stellar magnitudes above
the saturation limit.
The dashed line in Figure~\ref{fig:PTF_Mag_RMS} represents the
level of the measured asymptotic RMS for this case.
Note that, unless indicated explicitly, all the RMS values used are simple
standard deviations,
and not robust estimators.

\begin{figure}
\centerline{\includegraphics[width=8cm]{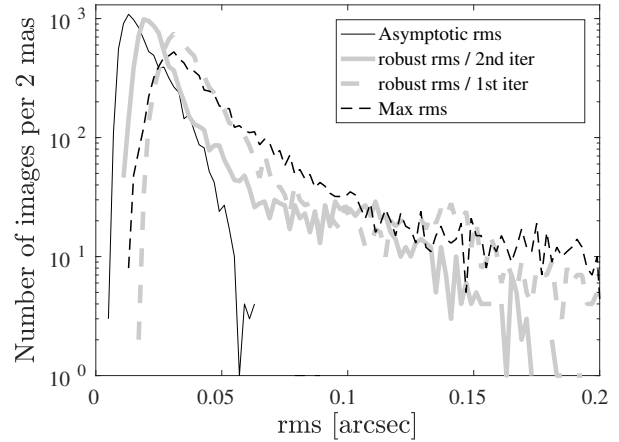}}
\caption{Distributions of RMS measurements
over the sample of PTF images.
The solid-black line shows the two-axes combined asymptotic RMS, that peaks at 13.8\,mas.
The solid gray line presents the total robust RMS
of all the stars used in the solution (i.e., not only the bright stars),
peaking at 21.3\,mas.
The dashed-gray line shows the robust RMS of all the matched stars
in the images (i.e., before the sigma clipping),
peaking at 31.3\,mas.
The dashed black line shows the distribution of RMS of all stars used in
the astrometry for the 512$\times$512 pixels sub image which has the largest RMS
in each image (peaking at 31.3\,mas).
The single-axis RMS values are typically $\sqrt{2}$ times smaller
than the combined two-axes RMS.
\label{fig:PTF_RMS}}
\end{figure}

Figure~\ref{fig:PTF_RMS} shows the distribution of RMS measurements
over the sample of PTF images.
The solid black line represents the asymptotic RMS.
This curve peaks at 13.8\,mas,
and its lower and upper 5-percentile are at 8.9\,mas and 38.9\,mas, respectively.
The worst asymptotic RMS
I encountered in this sample is 89\,mas.
On closer inspection of the four
images with asymptotic RMS larger than $80$\,mas,
I found that three of them have poor seeing ($>6.9''$) and
the fourth has a double-peaked PSF due to a tracking problem.
The solid gray line presents the total robust RMS
of all the stars used in the solution (i.e., not only the bright stars).
As expected, due to the Poisson noise, it has a longer tail.
Also shown as dashed gray line, is the robust RMS of all the matched stars
in the image (i.e., before the sigma clipping).

The solutions of the ZTF images have RMS values that are about $\sqrt{2}$
larger than those in the PTF images.
This is presumably due the fact that the ZTF integration time
is half of the PTF exposure times, and that the
astrometric scintillation noise decreases like the square root
of the integration time (e.g., Shao \& Colavita 1992).

\begin{figure}
\centerline{\includegraphics[width=8cm]{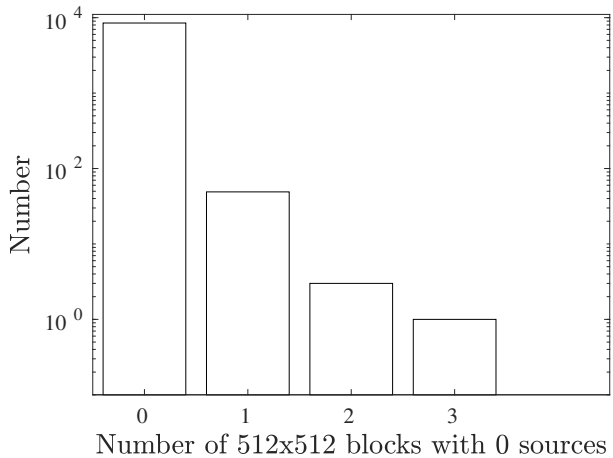}}
\caption{Histogram of the number of images in which, 0, 1, 2, and 3
blocks (i.e., $1024\times1024$\,pix sub images in each image)
have zero matched stars.
\label{fig:PTF_N0}}
\end{figure}

An important test is to verify that the solution is good over
the entire image.
In order to test this, the code reports the number of matches and RMS
in each sub-image of default size of $512\times512$ pixels.
Figure~\ref{fig:PTF_N0} presents a histogram of the
number of images in which, 0, 1, 2, and 3
blocks (i.e., $1024\times1024$\,pix sub images in each image)
have zero matched stars.
I found that 49 images had one $512\times512$-pixels
sub-image with zero matches, three images had
two sub-images with zero matches,
and one image had
three sub-images with zero matches.
Zero matches in a small sub-image
typically happens when the total number
of stars in the image is low and/or in regions with
elevated background.
I inspected by eye all the cases with more than one
sub-image with zero matches
to verify the solutions are correct.

\begin{figure}
\centerline{\includegraphics[width=8cm]{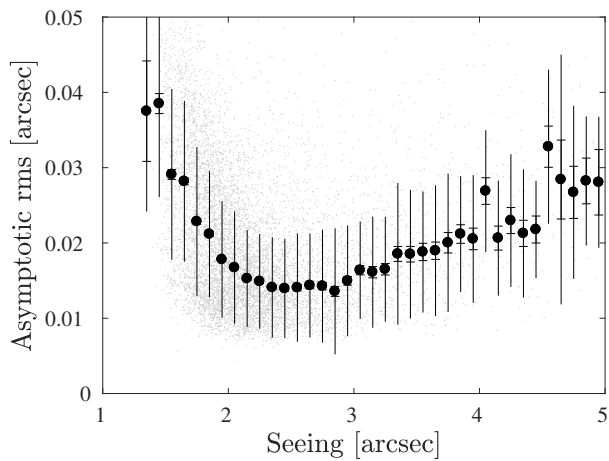}}
\caption{The two-axes combined asymptotic RMS as a function of seeing.
The gray points show the individual measurements,
while the black circles indicate the median asymptotic RMS
in each bin.
The full error bars represent
the standard deviation in each bin, while the horizontal bars indicate
the errors on the mean as estimated by the standard deviation
divided by the square root of the number of points in the bin.
\label{fig:PTF_Seeing_RMS}}
\end{figure}

Another important property, which is relevant for
all astrometric solvers, is the dependence of
the quality of the solution on airmass and seeing.
Figure~\ref{fig:PTF_Seeing_RMS} presents the asymptotic RMS as
a function of the seeing.
The gray points show the individual measurements,
while the black circles indicate the median asymptotic RMS
in seeing-bins.
The full error bars represent
the scatter of the points in each bin as calculated using
the standard deviation, while the horizontal bars indicate
the errors on the mean as estimated by the standard deviation
divided by the square root of the number of points in the bin.
The plot shows a clear minimum in the RMS around
a seeing of $\approx2.5''$.
The degradation in the quality of the solution below
seeing of about $2''$ is presumably due to under-sampling
of the PSF.
The PTF pixel scale is $1.01''$\,pix$^{-1}$.
Although the Nyquist sampling criterion
is not well defined for a Gaussian PSF\footnote{The Fourier transform of a Gaussian is another Gaussian and, therefore, is not bandlimted. However, in practice, the information content in the edges of the Gaussian is small.},
it is roughly requires that the full width at half maximum (FWHM) of the PSF
contain about two pixels.
Therefore, when the seeing is better than $\ltorder2''$,
the accuracy at which the source position
can be determined degrades.
For seeing above $2''$, the RMS increases again.
Fitting a power-law to the RMS vs. seeing,
in the range of $3''$ to $4.5''$ yielded
a power-law index of $1.04\pm0.15$.

\begin{figure}
\centerline{\includegraphics[width=8cm]{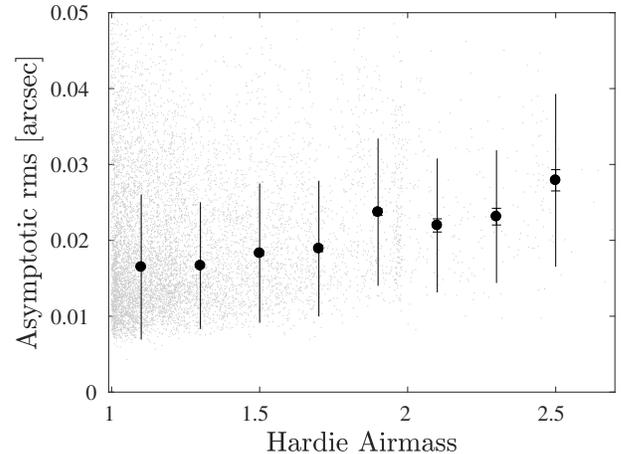}}
\caption{The asymptotic RMS
as a function of the Hardie airmass
(symbols like in Figure~\ref{fig:PTF_Seeing_RMS}).
\label{fig:PTF_AM_RMS}}
\end{figure}

Figure~\ref{fig:PTF_AM_RMS} shows the asymptotic RMS
as a function of the Hardie airmass
(symbols like in Figure~\ref{fig:PTF_Seeing_RMS}).
%The RMS degrade with airmass presumably due to
%the degradation in seeing.
%
The best fit power-law between the RMS and airmass,
shows that the RMS goes like the airmass to the power
of $0.54\pm0.09$.
The degradation of astrometry with
airmass is expected due to several reasons.
First, the seeing depends on the airmass.
Indeed, in these PTF images, the Spearman rank correlation
between the airmass and the seeing is 0.16
(with a false alarm probability $\ltorder10^{-5}$).
An additional reason is that 
the atmospheric color-dependent refraction
increases with airmass, which will
tend to increase the RMS as a function of airmass
regardless of the seeing (see example in Zackay, Ofek, \& Gal-Yam 2016).

I also tested the convergence of the astrometric accuracy
as a function of the transformation order.
For example, for one PTF image, I found that the asymptotic
RMS was $0.1772''$, $0.0504''$, $0.0288''$
and $0.0276''$ for
an affine transformation,
an affine transformation plus tip/tilt terms,
an affine transformation plus tip/tilt terms plus third-order Chebyshev polynomials of the second kind,
and an affine transformation plus tip/tilt terms plus fourth-order Chebyshev polynomials of the second kind, respectively.

\section{Code}
\label{sec:code}

The code is available\footnote{https://webhome.weizmann.ac.il/home/eofek/matlab/}$^{,}$\footnote{https://github.com/EranOfek/MAAT}
as part of the MATLAB
Astronomy \& Astrophysics Toolbox (Ofek 2014).
The main input arguments of the code, as well as usage examples,
are provided in the online
documentation\footnote{https://webhome.weizmann.ac.il/home/eofek/matlab/doc/astrometry.html}.

The code was developed using an object-oriented approach.
Specifically, the images are stored in a container of
astronomical images ({\tt SIM} class\footnote{https://webhome.weizmann.ac.il/home/eofek/matlab/doc/SIM.html}).
Each container (object) optionally contains a background, noise,
weight and bit-mask images, as well as a PSF, source catalog
and header information.
Each method (i.e., a function that operates on a class),
including the main code described here,
may create, use, and update the associated metadata.
For example, if one method populates the catalog field
in the image object, another method can use it, or modify it.
In this case, a single object represents many types of data
associated with an image.
One outcome is that the image is always associated with its metadata
(e.g., mask, background, catalog),
and the user does~not need to be concerned with
how to access and update the metadata.
Another advantage is that the {\tt SIM} object can be
an array of images and their metadata stored in memory.
This may alleviate the need for many time-consuming and unnecessary
read/write-to-disk operations.

\section{Summary}
\label{sec:Sum}

I present a new code for solving the astrometry of astronomical images.
The main feature of the code is its robustness against failures,
which I demonstrate on a large number of wide-field images.
The RMS of the astrometric residuals,
obtained using this code, are presumably close
to the atmospheric scintillation limit.
This robustness is mainly achived by three steps:
(1) removing sources found in regions of high source density;
(2) Partitioning the image to sub-images in which each region
is solved autonomously, and than using all the sources found
in sub-images with good solutions for a global fit;
and (3) Using orthogonal polynomials.

As a future prospect,
a possible method to further improve the astrometric solutions
is to utilize the fact that the astrometric scintillation
noise is highly correlated
(at least on arcmin-scales; e.g., Shao \& Colavita 1992)
with a correlation function that drops like
\begin{equation}
\sigma^{2}\propto\theta^{2/3},
\label{eq:Corr}
\end{equation}
where $\sigma^{2}$ is the variance and $\theta$ is the angular distance.
Using this correlation function for the covariance
of the fitted transformation,
along with higher order polynomials,
has the potential to reduce the residuals.
This approach may be especially important
in high stellar density fields.
Another important task is to reduce the systematic errors due to the instrument
(e.g., variations in the detector pixel size).
Additional planned improvements are adding color terms
to the transformation,
and trying to identify examples of failures
and fixing the code to solve these cases.

\acknowledgments

I would like to thank Peter Nugent and an anonymous referee for comments on the manuscript.
I am grateful for support by
grants from the Israeli Ministry of Science,
ISF, Minerva, BSF, BSF transformative program,
Weizmann-UK, and the I-CORE Program of the Planning
and Budgeting Committee and the Israel Science Foundation (grant No 1829/12).
Based on observations obtained with the Samuel Oschin Telescope 48-inch and the 60-inch Telescope at the Palomar Observatory as part of the Zwicky Transient Facility project. Major funding has been provided by the U.S National Science Foundation under Grant No. AST-1440341 and by the ZTF partner institutions: the California Institute of Technology, the Oskar Klein Centre, the Weizmann Institute of Science, the University of Maryland, the University of Washington, Deutsches Elektronen-Synchrotron, the University of Wisconsin-Milwaukee, and the TANGO Program of the University System of Taiwan.

\end{document}